\documentclass[12pt]{article}
\usepackage{amsmath}
\usepackage{amsfonts}
\usepackage{mathrsfs}
\usepackage{graphicx}

\newcommand{\nc}{\newcommand}
\newcommand{\rnc}{\renewcommand}
\nc{\bea}{\begin{eqnarray}}
\nc{\eea}{\end{eqnarray}}
\nc{\be}{\bea}
\nc{\ee}{\eea}
\nc{\trac}[2]{{\textstyle\frac{#1}{#2}}}
\nc{\del}{\partial}
\rnc{\d}{\delta}
\nc{\ra}{\rightarrow}

\title{}
\date{}
\begin{document}
\thispagestyle{empty}

\vspace*{2cm}
\begin{center}
{\Large\sc The non-Abelian gauge theory of\\ Matrix Big Bangs}
\end{center}
\vspace{0.2cm}

\begin{center}
{\large\sc Martin O'Loughlin${}^{a,}$\footnote{martin.oloughlin@ung.si}}\textsc{and} 
{\large\sc Lorenzo Seri${}^{b,}$\footnote{seri@sissa.it}}\\[.8cm]
{\it ${}^a$ University of Nova Gorica, Vipavska 13, 
5000 Nova Gorica, Slovenia}\\[.3cm]
{\it ${}^b$ SISSA, Via Beirut 3, 34151, Trieste, Italy and\\ 
INFN, Sezione di Trieste, Via Valerio 2, 34127, Trieste, Italy}
\end{center}

\vspace{1cm}
\noindent
We study at the classical and quantum mechanical level the
time-dependent Yang-Mills theory that one obtains via the generalisation
of discrete light-cone quantisation to singular homogeneous
plane waves. The non-Abelian nature of this theory
is known to be important for physics near the
singularity, at least as far as the number of degrees of freedom is
concerned. We will show that the quartic interaction is
always subleading as one approaches the singularity and that close
enough to $t=0$ the evolution is driven by the diverging tachyonic
mass term. The evolution towards asymptotically flat space-time also
reveals some surprising features.

%

\newpage
\section{Introduction}
The presence of singularities in the classical solutions to the
Einstein equations, or to extensions of these equations that include
various additional matter fields, is by now a well-established fact,
initially proven for the Einstein equations by Hawking and Penrose
\cite{Hawking:1969sw}. In general singularities are
regions of space-time where there is some breakdown in our classical
description of the physics. Furthermore, even if we consider a
singular metric configuration corresponding to a vacuum, 
i.e. pure gravity with no matter, small perturbations due to the 
introduction of matter stress-energy or arising from quantum fluctuations 
of matter fields will inevitably produce 
divergences in the equations that describe the further evolution of the 
system. To make further progress in the understanding of singularities one 
needs some theory of quantum gravity.

String Theory is a promising candidate for a theory of quantum gravity
and one of our main purposes in this paper is to show how models of
singularities constructed from string theory can lead to new
insights. Indeed it has been shown that string theory can ``resolve''
certain kinds of time-independent singularities (euclidean orbifolds
\cite{Dixon:1985jw, Dixon:1986jc}). Cornalba and Costa
\cite{Cornalba:2002fi} and Liu, Moore and Seiberg \cite{Liu:2002ft,
Liu:2002kb} attempted to extend these results to singularities that
display some features of a Big Bang (time-dependent orbifolds),
investigating in particular whether it is possible to define a
temporal evolution through a singularity. This analysis led to
divergent amplitudes and furthermore the actual stability of the
backgrounds used in these works was brought into question
\cite{Horowitz:2002mw}. Moreover these backgrounds are almost
everywhere flat, being built from Minkowski space via orbifolding,
thus making them unsuitable for modelling real cosmological solutions
that possess non-trivial curvature.

More recently an alternative approach to the study of cosmological
singularities has been introduced by Craps, Sethi and Verlinde
\cite{Craps:2005wd} where they carry out a Discrete Light-Cone
Quantisation (DLCQ) on a type-IIA background given by a flat
Minkowskian metric and a dilaton linear in time. In
\cite{Susskind:1997cw} it is conjectured that DLCQ in Minkowski space
enables us to non-perturbatively describe a certain sector of string
theory, singled out by fixing the light-cone momentum. The CSV
generalisation of this procedure, which simply involves the addition
of a linear dilaton to the Minkowski background, leads to a
1+1-dimensional SYM theory with a time-dependent coupling constant
$g_{\mathrm{YM}}\sim 1/g_\mathrm{s}\sim e^{ct}$ that is inversely
proportional to the string coupling (see also \cite{Craps:2006yb} for
a clear review of this construction). In other relevant work, as presented in 
\cite{Craps:2007ch} and references therein, an alternative and promising
approach to the resolution of cosmological singularities that uses the
AdS/CFT correspondence is pursued.

As mentioned, the CSV matrix big bang is derived using a flat
background. It was subsequently demonstrated in \cite{Blau:2008bp}
that a further generalisation is possible such that the DLCQ can be
applied to an entire class of metrics, the Singular Homogeneous Plane
Waves (SHPWs), that are non-trivially curved and singular along a null
submanifold. Furthermore, these metrics are of relevance to the physics of
space-time singularities as they can be found as the Penrose Limits of a
large variety of singular metrics, including the Schwarzschild black
hole and the Friedmann-Robertson-Walker universe \cite{Blau:2004yi}
(time-dependent plane-waves have also been used as backgrounds for
string theory, see \cite{Papadopoulos:2002bg}). The final theory that
one obtains is again a 1+1-dimensional SYM theory (given in equation
\eqref{sbc} of section 2) where the coupling constant now depends on
time as a power $t^q$ and is again inversely proportional to
$g_\mathrm{s}$.  The power $q$ may be positive or negative but for us
the interesting cases correspond to $q>0$ for which the Yang-Mills
coupling constant becomes small near the singularity (namely $t=0$)
and blows up for large times, exactly as it does in the CSV model.

We are interested in studying the two
limiting cases of $t$ close to zero and $t$ very large. It is believed 
that in these two limiting regions the following occur:
\begin{itemize}
\item
near the
singularity the smallness of $g_{YM}$ allows the fields (coordinates)
to be non-commuting and more degrees of freedom will become available. 
The singularity would then be in a region where the
geometry of the space time is intrinsically non-commutative\footnote{As has also been demonstrated in the case of D-instantons near the 
singularity of the Misner Universe \cite{She:2005mt}.}.
This is 
indeed one of the general messages of CSV \cite{Craps:2005wd} and in 
the present article we present various pieces of evidence to further
support this picture in the more general context of SHPWs;
\item
for large times the quartic interaction term built up from the
product of two commutators is forced to zero in order to contrast the
growth of the time-dependent $g_{YM}$, implying that the fields become
ordinary commuting coordinates (the matrices are in the Cartan
subalgebra) and the model reduces to the light-cone quantisation of 
string field theory in Minkowski space-time with a linear dilaton background.
Again this is one of the messages of the CSV paper in for which, 
however, we encounter some difficulty in obtaining further support.
\end{itemize}

Our main goal in this paper is to investigate the validity of these
qualitative conclusions. We will study a simple non-Abelian
time-dependent theory that is suitable for modelling the
1+1-dimensional SYM theory obtained via the DLCQ of SHPWs. We discuss
the relevance of the non-Abelian interaction near the singularity both
at the classical and at the quantum mechanical levels and the way in
which the normal (commutative) space-time physics may emerge from the
non-Abelian physics that appears to be important near the
singularity. Our paper readdresses some of the issues discussed in
\cite{Awad:2008jf,Craps:2008bv} in particular as regards the physics
of the singularity and the role of the quartic interaction term. We
have some new insights into these questions found by studying a toy
model that retains more of the non-Abelian physics in comparison to
the one-dimensional model discussed in \cite{Awad:2008jf}.

The contents of the paper are as follows: in Section 2 we review DLCQ
on SHPWs, and we present the toy model that we are going to study; in
Section 3 we perform a classical analysis of this model, discussing
both analytically and numerically the way in which the non-Abelian
interaction near $t=0$ can be considered as a perturbation and also
making some considerations regarding the $t\rightarrow\infty$ physics.  In
the following section we begin a discussion of the quantum mechanics
of our toy model, studying in subsection 4.1 the Lagrangian without
quartic interaction. In subsections 4.2 and 4.3 we deal with the
quantum generalisation of the classical perturbative analysis of
section 3. We explain why the naive perturbative approach fails in
certain cases and discuss the conditions for which the non-Abelian
interaction can be said nonetheless to be always subleading near the
singularity. The final subsection of section 4 contains a short
discussion and interpretation of the quantum mechanics in the limit
that $t\to\infty$.  Section 5 contains a summary of the results and
some conclusions.

\section{DLCQ and non-Abelian Yang-Mills}

The basic setup that we will be using throughout this paper is that of the
DLCQ of SHPW's as presented in \cite{Blau:2008bp}. Before we begin we
will provide a brief summary of the relevant results of that paper. 

Homogeneous Plane Waves are particular plane wave space-times that are
homogeneous, meaning that they have a transitive isometry algebra
\cite{Blau:2002js}. The generic plane wave metric in Brinkmann
coordinates is
\begin{equation}
ds^2 = - 2 dx^+dx^- + A_{ij}(x^+)x^ix^j(dx^+)^2 + (d\vec{x})^2.
\end{equation}
This space has $2d-3$ Killing vectors corresponding to a Heisenberg algebra
of $2(d-2)$ components and an extra translation Killing vector $\partial_-$. 
Homogeneous plane waves have an additional Killing vector related to 
a transformation in the $x^+$-direction. For (non-singular) 
homogeneous plane waves this arises when $A_{ij}(x^+)$ is constant in which case
the additional Killing vector is related to translations in $x^+$. Singular
homogeneous plane waves instead correspond to $A_{ij}(x^+) = B_{ij}/(x^+)^2$ in 
which case the additional Killing vector is related to the scaling 
$(x^+,x^-)\rightarrow (\mu x^+,\mu^{-1} x^-)$. 
For SHPWs the DLCQ prescription involves three steps, each related to
symmetries of the space-time. First the alignment (by a null rotational
isometry) of an almost null-circle with a space-like circle, then
a rescaling of all energies by a boost isometry that acts directly on
$\partial_+$, and finally a rescaling of the coordinates
(mass/length scales) as originally proposed by Seiberg and
Sen \cite{Seiberg:1997ad,Sen:1997we} using a homothety of the metric.

In this paper we will restrict attention to the class of SHPW metrics
that arise when taking the Penrose limit of space-time singularities 
\cite{Blau:2004yi}. In Brinkmann coordinates these are
\be\label{3plrc}
ds^2 = -2 dx^+dx^-  + \sum_i \frac{p_i(p_i+2)}{4(x^+)^2}(x^i)^2(dx^+)^2 
+\sum_i(dx^i)^2.
\ee
We will be interested in solutions to the type IIA string theory 
with a dilaton of power law form,
\begin{equation}
\mbox{e}^\phi=(x^+)^{-q},
\end{equation}
as dictated by the choice of metric and equations of motion to be discussed
below. 

Applying the DLCQ procedure to this system of metric and dilaton one
finds a two-dimensional matrix string theory described by the action,
\begin{multline}\label{sbc}
S = \frac{1}{2}\int d^2\sigma\;\mathtt{Tr}\left( -\frac{1}{2g^{2}_{YM}(t)}
\eta^{\alpha\gamma}\eta^{\beta\delta}F_{\alpha\beta}F_{\gamma\delta}
-\eta^{\alpha\beta}\d_{ij}D_{\alpha}X^i D_{\beta}X^j\right.\\
+ \left.g^2_{YM}(t) \d_{ik}\d_{jl}[X^i,X^j][X^k,X^l] +
\sum_i \frac{p_i(p_i+2)}{4t^2}(X^i)^2 \right)\;\;,
\end{multline}
where we are in light-cone gauge $x^+=t$ and the coupling constant 
of the gauge theory is related in a simple manner to the original dilaton,
\be
\label{gym}
g_{YM}(t) \sim \frac{1}{g_s\ell_s} \mbox{e}^{-\phi}= \frac{1}{g_s\ell_s} t^q\;\;.  
\ee 
This construction provides us, in principle, with a non-perturbative
description of the full string theory in a sector of fixed light-cone
momentum.  Note the simplicity of this final result. After a series of
T and S-dualities on the original metric and dilaton of type IIA
string theory, we find this matrix string action again containing the
original type IIA metric. Note also that it is only with this choice
of Brinkmann coordinates that the metric contains information free of
ambiguities. A study of the same system using Rosen coordinates is
likely to be fraught with interpretational difficulties due to the
fact that these coordinates do not fix a unique form for the metric
and thus generally contain extraneous non-geometric information.

The equations of motion of 
type IIA string theory give a precise relationship between the frequencies
determined by the $p_i$ and the behaviour of the dilaton,
\begin{equation}
\frac{1}{4}\sum_i p_i(p_i+2) = 2q. 
\end{equation}
In this paper we will consider primarily two particular relations
between $p$ and $q$ although much of the analysis and conclusions are 
independent of these choices. Using the above equation of
motion for string theory on a SHPW with d-coordinates compactified on
torii and the $8-d$ remaining  $p_i=p$ equal we find
\begin{equation}
q=(8-d)p(p+2)/8.
\end{equation}
On the other hand, for the
purpose of comparison between our results and those of
\cite{Awad:2008jf} we also consider the case $p=2q$ which arises for 
type IIB D3-branes as constructed in \cite{Awad:2007fj}.

We are only interested in the case in which $q>0$ as it is in this
case that the string coupling constant diverges and the Yang-Mills
coupling goes to zero like $|t|^q$ as the singularity at $t=0$ is
approached. We see immediately that for the same range of $q$ and
$|t|\rightarrow\infty$ we recover an asymptotically flat space-time
with weak string coupling and a diverging Yang-Mills coupling. This
means in particular that the coefficient of the quartic commutator
term in the potential diverges, presumably fixing the transverse
coordinates to be mutually commuting in the limit that the space-time
is asymptotically (light-cone) flat. In the following we will discover
that for the time-dependent system the situation is not so simple.

\subsection{The toy model}

To make a quantitative study of this system we will first compactify
the action \eqref{sbc} 
on a world-sheet circle of radius $2\pi L$ and then expand
it as a sum over the modes of the fields. Each of these modes is then
described by a quantum mechanical system with a common time-dependent
frequency and a quartic coupling,
\begin{multline}
S = \frac{1}{2}\int dt\;\mathtt{Tr}\left(\sum_n 
\dot{X}^{i\dagger}_n\dot{X}^i_{-n} 
- \omega_n(t)^2 X^{i\dagger}_nX^i_{-n}\right. \\
\left.+ \lambda\sum_{n_1,n_2,n_3}
|t|^{2q}[X^{i\dagger}_{n_1},X_{n_2}^{j\dagger}]
[X^k_{n_3},X_{-n_1-n_2-n_3}^l]\delta_{ik}\delta_{jl}\right)
\end{multline}
where the time-dependent frequency is
\begin{equation}\label{time-dep-freq}
\omega_n(t)^2 = \frac{n^2}{2\pi L} - \frac{p(p+2)}{4t^2}=
\omega_n^2 - \frac{p(p+2)}{4t^2}\;.
\end{equation}
To reduce this still quite complicated and non-linear system to something 
more tractable we will consider just two
$SU(2)$ matrices corresponding to two of the eight transverse
coordinates and we will focus attention on a single KK mode of each
coordinate with action
\begin{equation}\label{stoy}
S_\mathtt{toy} = \frac{1}{2}\int\!dt\;\mathtt{Tr} \bigl(|\dot{X}|^2 
+ |\dot{Y}|^2 - \omega_X(t)^2 |X|^2 -
\omega_Y(t)^2|Y|^2  + \lambda|t|^q|[X,Y]|^2\bigr)
\end{equation}
where
\begin{equation}
\omega_{X,Y}(t)^2 = \omega_{n_X,n_Y}^2 - \frac{p(p+2)}{4t^2}\;.
\end{equation}
We believe that this is the simplest model that retains the important
features of the full non-Abelian Yang-Mills theory (compare with
\cite{Awad:2008jf}). It has a time-dependent mass term and a
time-dependent quartic potential representative of the interaction
term of non-Abelian gauge theory.

Note that for the case of interest to us $q>0$ and thus the harmonic
oscillator frequencies become purely imaginary as $|t|$ approaches
zero. This fact will play an important role in the study of the
non-Abelian physics of the singularity.  This relation between
gravitational singularities at strong string coupling and tachyonic
behaviour in the DLCQ matrix theory is more general than that
mentioned here and further discussion of this point can be found
in \cite{Blau:2008bp}.

A time-independent theory with the same Lagrangian but constant
frequencies and a constant quartic coupling is well studied and it has
been shown that if the energy is large with respect to the curvature
of the potential then the quartic potential leads to a chaotic behaviour of
the scalar fields \cite{Aref'eva:1998mk}. It is further believed that
if the energy is small compared to the curvature of the potential and
if the quadratic term is exactly zero (as it would be for a
supersymmetric system) then when the tension is taken to be very large
the system is forced into a configuration consisting of commuting
matrices \cite{Witten:1995im}.

In the limit that $t\rightarrow 0$ our non-linear problem becomes a
time-dependent harmonic oscillator with a quartic perturbation.  We
are most interested in determining to what extent this quartic term
can truly be treated as a perturbation. As the behaviour of this
time-dependent system is non-trivial  we will first take a
look at the classical physics, both analytically and
numerically. Secondly we review the quantum mechanics in the inverted
time-dependent oscillator and then attempt to extend this analysis to
the full quantum mechanical problem using time-dependent perturbation
theory. This analysis will lead us to conclude that near the
singularity the quartic non-Abelian interaction is not important
whereas at infinity it plays an important role in the evolution of the
system.

\section{The classical physics}

First of all, let us solve the classical equations for the time-dependent
harmonic oscillator (taking $\lambda=0$),
\begin{equation} \label{modeeq}
\ddot{X}+\omega(t)^2X=\frac{\mathrm{d}^2}{\mathrm{d}t^2}X-
\frac{p(p+2)}{4t^2}X+\omega^2X=0.
\end{equation}
It is straightforward to see that one can equivalently write
\begin{equation}
\frac{\mathrm{d}^2}{\mathrm{d}t^2}\left(\frac{X}{\sqrt{t}}\right)
+\frac{1}{t}\frac{\mathrm{d}}{\mathrm{d}t}\left(\frac{X}{\sqrt{t}}\right)
+\left(\omega^2-\frac{(p+1)^2}{4t^2}\right)\frac{X}{\sqrt{t}}=0,
\end{equation}
which is Bessel's equation for $X/\sqrt{t}$. The solutions are, 
\begin{equation}
X(t)=\sqrt{\omega t}\left\{a H_\nu(\omega t)+a^* H^*_\nu(\omega t)\right\},
\end{equation}
where $H_\nu(\omega t) = J_\nu(\omega t) + i Y_\nu(\omega t)$, 
$J_\nu$ and $Y_\nu$ are Bessel functions of the first and second kind and
$\nu\equiv \frac{p+1}{2}$.

\subsection{Perturbation theory}

Now we rewrite our interaction term using the properties of 
the Pauli matrices of $SU(2)$:
\begin{equation}
\lambda |t|^{2q} \mathtt{Tr}|[X,Y]|^2 = 
\lambda |t|^{2q}\left(|\vec{X}|^2||\vec{Y}|^2-(\vec{X}\cdot\vec{Y})^2\right)
\end{equation}
The equations of motion are
\begin{equation}
\begin{split}
\hat{L}(t)X_i+\omega_{n_X}^2X_i&=\lambda |t|^{2q}
\left(2X_i|\vec{Y}|^2-2Y_i\vec{X}\cdot\vec{Y}\right),\\
\hat{L}(t)Y_i+\omega_{n_Y}^2Y_i&=\lambda |t|^{2q}
\left(2Y_i|\vec{X}|^2-2X_i\vec{X}\cdot\vec{Y}\right),
\end{split}
\end{equation}
where the operator $\hat{L}(t)= -\frac{\partial^2}{\partial t^2} +
\frac{p(p+2)}{4t^2}$.  We want to study the possibility of treating the
quartic term as a perturbation and so we will expand around the
solutions to the time-dependent harmonic oscillator ($\lambda=0$),
choosing for simplicity the initial conditions $X_{1,2}=0$ and $\dot{X}_{1,2}=0$. 

For the (hopefully small) perturbations we find the linearized equations
\begin{equation}\label{pert1}
\begin{split}
\hat{L}(t)\delta X_i+\omega_{n_X}^2\delta X_i & =\lambda |t|^{2q} \left(2X_i\vec{Y}^2+4X_i\vec{Y}\cdot\delta\vec{Y}+2\vec{Y}^2\delta X_i-2 Y_i\vec{X}\cdot\vec{Y}\right.\\
& \quad \left.-2Y_i\vec{X}\cdot\delta \vec{Y}-2Y_i\vec{Y}\cdot\delta\vec{X}-2\delta Y_i\vec{X}\cdot\vec{Y}\right),\\
\hat{L}(t)\delta Y_i+\omega_{n_Y}^2\delta Y_i & =\lambda |t|^{2q} \left(2Y_i\vec{X}^2+4Y_i\vec{X}\cdot\delta\vec{X}+2\vec{X}^2\delta Y_i-2 X_i\vec{X}\cdot\vec{Y}\right. \\ 
& \quad \left.-2X_i\vec{Y}\cdot\delta \vec{X}-2X_i\vec{X}\cdot\delta\vec{Y}-2\delta X_i\vec{X}\cdot\vec{Y}\right).\\
\end{split}
\end{equation}
Inserting the unperturbed solution and keeping only the leading terms
on the rhs we then obtain, 
\begin{equation}\label{pert2} 
\begin{split}
\hat{L}(t)\delta
X_{1,2}+\omega_{n_X}^2\delta X_{1,2}& = \lambda 
|t|^{2q}\left(-2Y_{1,2}X_3Y_3\right),\\ 
\hat{L}(t)\delta
X_3+\omega_{n_X}^2\delta X_3& = \lambda
|t|^{2q}[2X_3(Y_1^2+Y_2^2)],\\ \hat{L}(t)\delta
Y_{1,2}+\omega_{n_Y}^2\delta Y_{1,2}& = \lambda |t|^{2q}
(2Y_{1,2}X_3^2),\\ \hat{L}(t)\delta Y_3+\omega_{n_Y}^2\delta Y_3 &= 0
\end{split}
\end{equation}
and we now want to specifically consider small $|t|$ where
there is a possibility that the rhs of these equations is truly a
perturbation to the harmonic oscillator. The generic solution to the
homogeneous unperturbed equation is, to leading order,
\[\mathrm{A}t^{-\frac{p}{2}}+\mathrm{B}t^{\frac{p}{2}+1}\]
for some constants $A$ and $B$. On the rhs of our equations 
\eqref{pert1}, \eqref{pert2} we always have the product of three 
solutions of this type and we can thus say that the only singular 
inhomogeneous term 
is of the form $Kt^{2q-3p/2}$. We have thus reduced our problem to 
that of solving (near $t=0$) equations of the general form
\begin{equation}
\left(\frac{\mathrm{d}^2}{\mathrm{d}t^2}-
\frac{p(p+2)}{4t^2}\right)F(t)=Kt^{2q-3p/2}\;.
\end{equation}
Using the ansatz $F(t) = Ct^\gamma$ we find
\begin{equation} 
\left\{
\begin{split}
\gamma =&2q -\frac{3p}{2}+2,\\
C=&-\frac{K}{P(p,q)},
\end{split}
\right.
\end{equation}
where $P(p,q)= 2p^2+ 4q^2 -6pq-5p+6q +2$. The perturbation theory is valid
when $2q>p-2$ and the solution is valid provided that $P(p,q)\neq0$.
In the range of $p,q$ for which perturbation theory is valid $P(p,q)=0$
when $p=q+1/2$ and in this special case one needs to replace the 
above ansatz by $F(t)\sim C t^\gamma log(t)$ finding
\begin{equation}
\left\{
\begin{split}
\gamma =&p/2+1,\\
C=&\frac{K}{2p+1}.
\end{split}
\right.
\end{equation}

We are interested in the behaviour of $\delta X\sim\delta Y\sim F(t)$
as $t\to 0$ and we see for the two special cases of $q=p/2$ and
$q=(8-d)p(p+2)/8$ that $F(t)$ is less singular than the unperturbed solution
as a consequence of choosing $2q-p+2>0$ and $q>0$. At the moment we
are implicitly considering initial conditions for which the classical
particle begins its motion with $X(t_0)$ near the origin and thus we
see that its evolution is predominantly determined by the harmonic
oscillator potential. In the next section we discuss the numerical
integration of the full non-linear equations thus allowing us to
consider initial conditions for which the quartic term is also
important. We will find that after an initial transient phase as $t$
approaches the singularity the particle will once again converge to a
solution determined entirely by the harmonic oscillator
potential. This is one of the key messages of this article and we will
return to discuss it several times in the following.

\subsection{Numerical integration of the toy model}

In addition to the qualitative study of these equations we were also
able to carry out their numerical integration with results that
largely corroborate the conclusions of the previous subsection.  There
are also some interesting surprises revealed by the numerical analysis
related to the role of the quartic interaction in the limit that
$|t|$ becomes large. Physically this is quite an interesting limit as
it should tell us something about the way that string theory in an
asymptotically flat space-time emerges from the non-Abelian physics 
near the singularity.

As we are interested in both large $|t|$, the asymptotically flat region, 
and small $|t|$, the singularity, we carried out numerical integration 
in both directions, beginning with some generic initial conditions at 
an intermediate value of $t$. 

\begin{figure}[h]
\centerline{\mbox{\includegraphics[width=4.5cm]{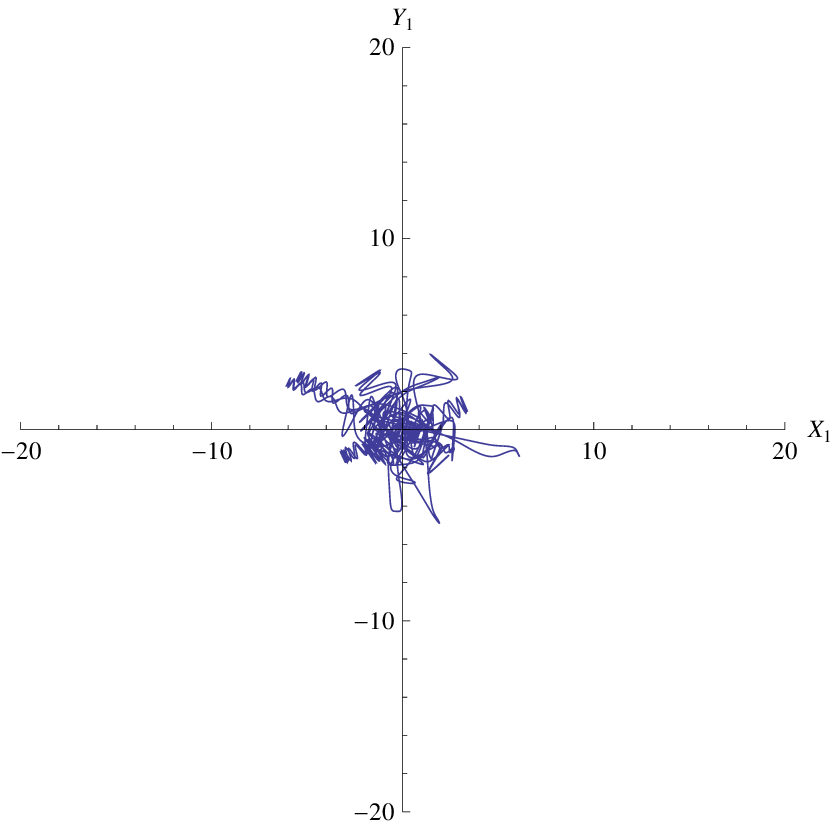}}
\mbox{\includegraphics[width=4.5cm]{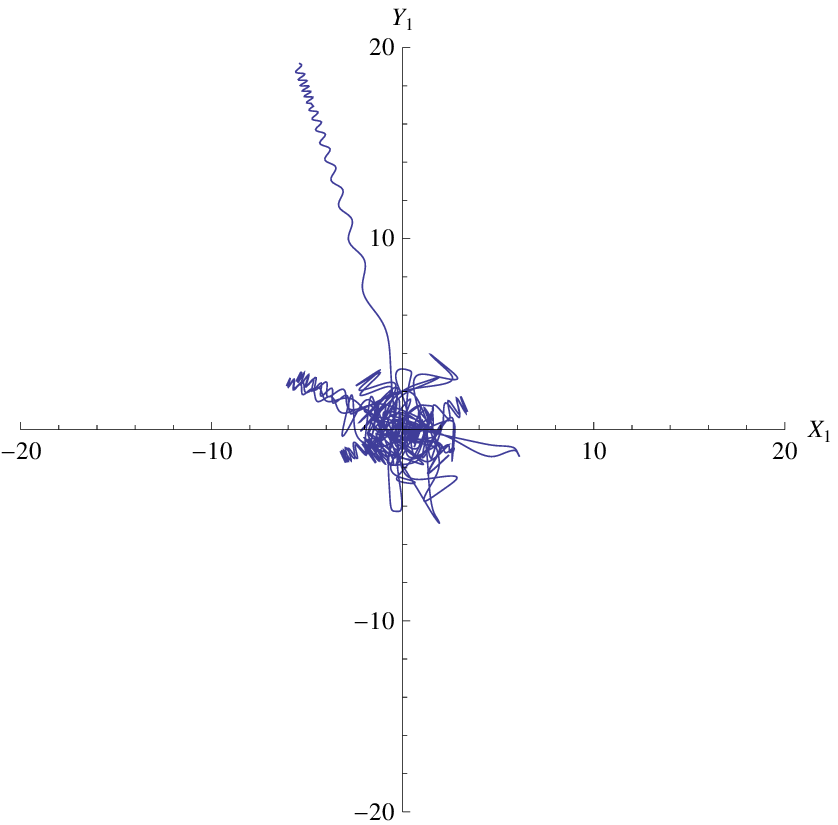}}
\mbox{\includegraphics[width=4.5cm]{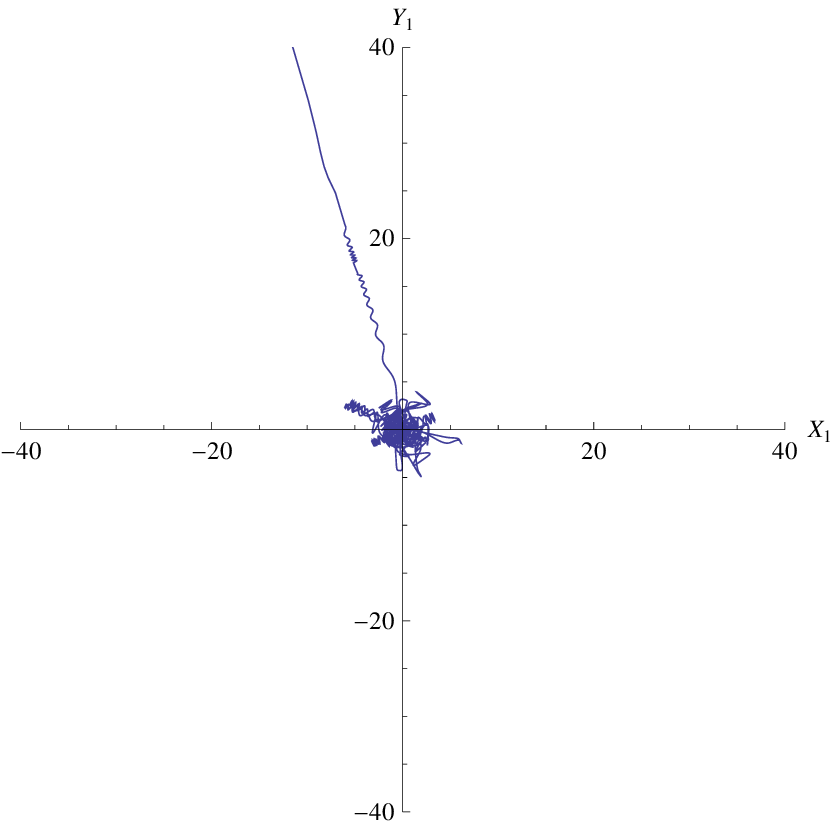}}}
\caption
{Numerical integration of toy model projected in the $X_1,Y_1$ plane
  for $t_i = -100$ and $t_f=-10,-1$ and $-.001$ respectively. Note the
  change in scale for the third plot to allow space for the runaway
  behaviour at $t\rightarrow 0$.}
\end{figure}

Imposing initial conditions at a sufficiently large negative $t$ and
studying the evolution as $t\rightarrow 0$ we find that the greater
part of the evolution is governed by the quartic interaction
term. This produces a generally oscillatory behaviour as the particle
moves down into a valley where the coordinates commute until at some
point, due to the steepening valley walls, the particle is forced to
turn around and leave the valley, then maybe bouncing around near
the origin or proceeding along another valley.  Once $t$ becomes
sufficiently small and $\omega_n(t)^2$ becomes
negative, the corresponding coordinate crosses over from this oscillatory
behaviour to a divergent tachyonic behaviour. This clearly indicates
that near $t=0$, for quite generic initial conditions, the classical
evolution is dominated by the quadratic $1/t^2$ terms in the
potential, as we also concluded in the previous subsection. Figure 1 is an 
example of a numerical integration displaying these features.

Studying more carefully the behaviour after the appearance of negative
curvature in the potential around $X=Y=0$ we see that there is a
transient phase of evolution during which the particle feels the
quartic interaction.  The duration of this transient depends on how
close the particle is to the minimum (the presence of which is a
consequence of the quartic potential) but universally, at the end of
this period and as $t$ gets closer to 0, the motion is dominated by
the divergent quadratic potential.

\begin{figure}[h]
\centerline{\mbox{\includegraphics[width=4.5cm]{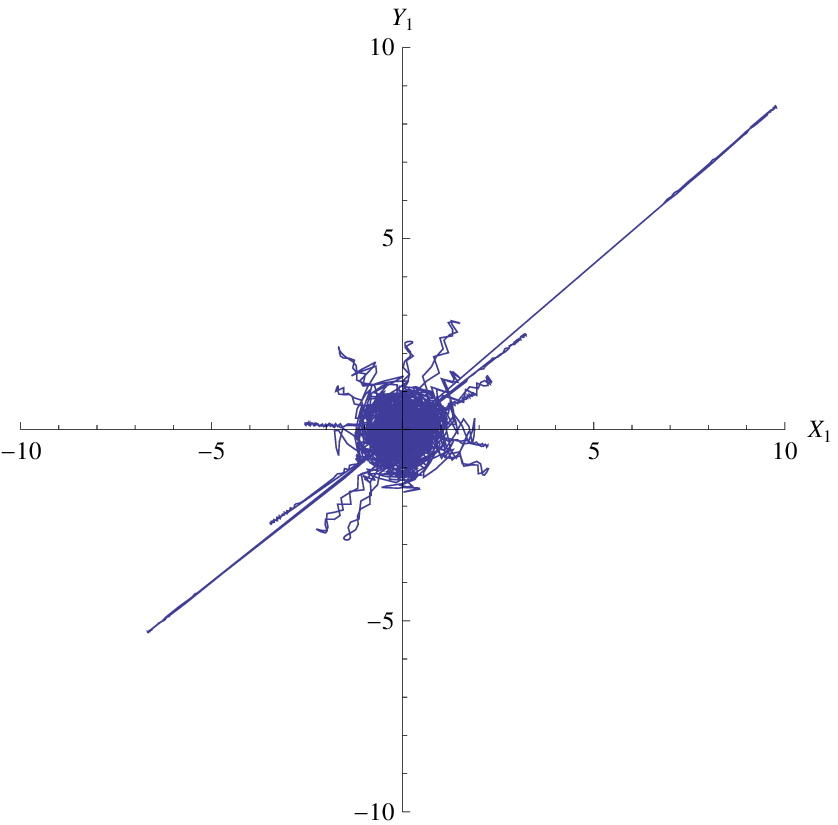}}
\mbox{\includegraphics[width=4.5cm]{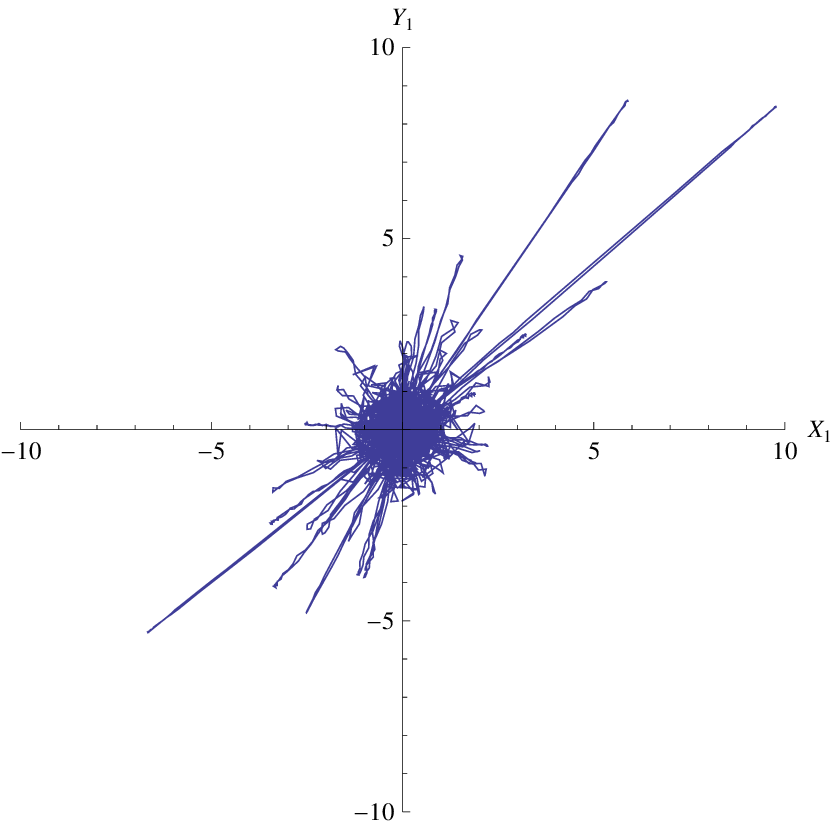}}
\mbox{\includegraphics[width=4.5cm]{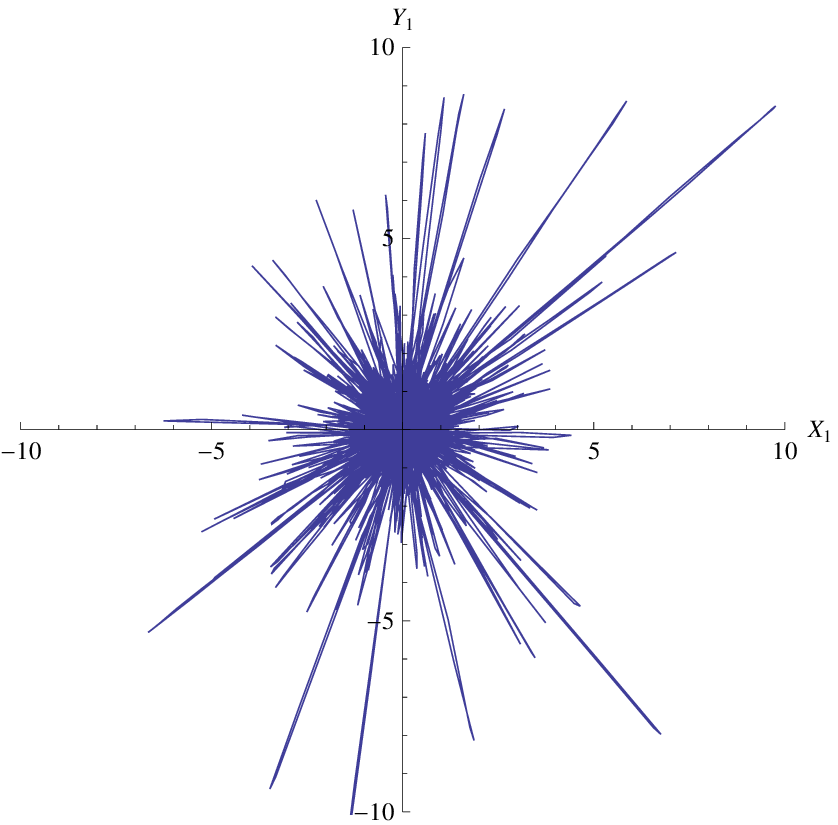}}}
\caption{Numerical integration of toy model projected in the $X_1,Y_1$ plane
for $t_i = 100$ and $t_f=500,1000$ and $5000$ respectively.}
\end{figure}

One may alternatively consider the singularity as the Penrose limit of an
initial cosmological singularity and impose initial conditions at small times
evolving the system to large positive times. Again, given an initial
$t$ small enough so that the tachyonic mass term dominates, the
evolution begins with a power law and divergent motion that then
settles down to oscillations driven by the quartic interaction. Once
again this term is not sufficient to restrict the matrices to commute
but rather their evolution appears to be chaotic, not settling down to
one direction in the Cartan subalgebra but randomly jumping from one
nearly commuting configuration to another. One can clearly see this
behaviour in Figure 2 where we have taken $p=0.5$ and $q=0.1$ and we
begin the integration in the oscillatory region. The increasing final
times $t_f$ illustrate the continual ``chaotic'' motion between
various nearly commuting configurations. The long spikes represent
periods during which the matrices are approximately commuting and we
see that the number of excursions into different commuting valleys
increases proportionally to the time elapsed. We will return to 
a discussion of the large $|t|$ behaviour in later sections of this paper.

Note that we have only considered the zero modes of the KK reduction
to quantum mechanics. Clearly the higher modes for large enough $t$
always have a positive harmonic oscillator potential and will be bound
to remain small and so the issue of commuting configurations does not
arise for them.

\subsection{Early times and robustness of the classical analysis: 
the $X^4$ case}

In order to check the validity of our analysis we further simplify to
the one-dimensional time-dependent anharmonic oscillator,
\begin{equation}\label{onedim}
S =\frac{1}{2}\int dt\bigl(\dot{X}^2-\omega(t)^2 X^2-\lambda|t|^{2q}
X^4\bigr).
\end{equation}
Though ill-suited for modelling the behaviour of our actual theory at
large times, the system given by \eqref{onedim} can provide us with
valuable information as far as small times are concerned. Indeed we
can think of the bidimensional potential $X^2Y^2$ as a special
configuration of our $SU(2)$ theory \eqref{stoy} 
(namely $X\perp Y$) in which case,
using the inequality $X^2Y^2 \leq \frac{1}{2}(X^4 + Y^4)$, we see that
the two decoupled copies of \eqref{onedim} provide us with an
overestimate of the potential in this configuration.

In particular, we would like to check that
\begin{itemize}
\item
as $t\to 0$ the solution found by including the quartic term is
increasingly well approximated by that for the tachyonic
$1/t^2$ oscillator;
\item
in the same limit the minimum of the potential runs off to infinity
faster than the maximum possible velocity of the classical particle so
that, after an initial transient evolution that depends critically on
initial conditions, the particle sees only the (divergent as $t\rightarrow 0$)
tachyonic mass.
\end{itemize}

Let us expand on this second point. At fixed $t$ the potential
shows a minimum located at position $X_{\mathrm{min}}(t)$ that receeds
to infinity as $t$ grows smaller.  In order to determine if a particle
can keep itself near $X_{\mathrm{min}}(t)$ during the complete evolution to
$t=0$ let us set the particle mass $m=1$ and observe the following bounds on
the time evolution: for fixed $t$ the maximal acceleration is seen to
be
\begin{equation}
a_{\mathrm{max}}(t)=\frac{\sqrt{2}[p(p+2)]^{3/2}}{12\sqrt{3\lambda}} 
|t|^{-q-3},
\end{equation}
and integrating we find (up to a constant term equal to the initial velocity)
an overestimate for the speed of a particle moving in this potential,
\[v_{\mathrm{max}}(t)=-\frac{\sqrt{2}[p(p+2)]^{3/2}}{12\sqrt{3\lambda}(q+2)} 
|t|^{-q-2} + c.\]
On the other hand the position of the minimum is
\begin{equation}
X_{\mathrm{min}}(t)=\frac{1}{|t|^{q+1}}\sqrt{\frac{p(p+2)}{2\lambda}},
\end{equation}
and it moves towards infinity with the speed
\begin{equation}
V_{\mathrm{min}}(t)=-\frac{q+1}{4}\sqrt{\frac{2p(p+2)}{\lambda}}|t|^{-q-2}.
\end{equation} 
The condition that for small enough $|t|$ the particle motion is not affected
by the quartic term in the potential is that 
\begin{equation}
|v_{\mathrm{max}}|<|V_{\mathrm{min}}|,
\end{equation}
and thus as the two velocities have identical $t$ dependence we simply need 
to compare their coefficients. This leads us to the inequality
\begin{equation}
3\sqrt{3}(q+1)(q+2)>p(p+2),
\end{equation}
and as we are assuming that $q>0$ this reduces to 
\begin{equation}
q>\frac{1}{2}\left(-3 + \sqrt{1 + 4\frac{p(p+2)}{3\sqrt{3}}}\right).
\end{equation}
For $0<p<-1 +\sqrt{1+6\sqrt{3}}$ this condition is satisfied without
further restricting the range of $q$ while for 
$p>-1+\sqrt{1+6\sqrt{3}}$ one needs $q> -3/2 + 1/2\sqrt{1 +
  4p(p+2)/3\sqrt{3}}$.  When $p=2q$ or $q=(8-d)p(p+2)/8$ there is no bound on
$p$ or $q$ apart from the already imposed requirement of positivity.

These general results can be easily checked in explicit examples by
numerically integrating the equations of motion for $X(t)$ and choosing various
initial conditions. As can be seen in figure 3 the particle after some initial 
transient oscillations is then quickly left behind by the minimum of the full
quadratic plus quartic potential.

\begin{figure}[h]
\centerline{\mbox{\includegraphics[width=6cm]{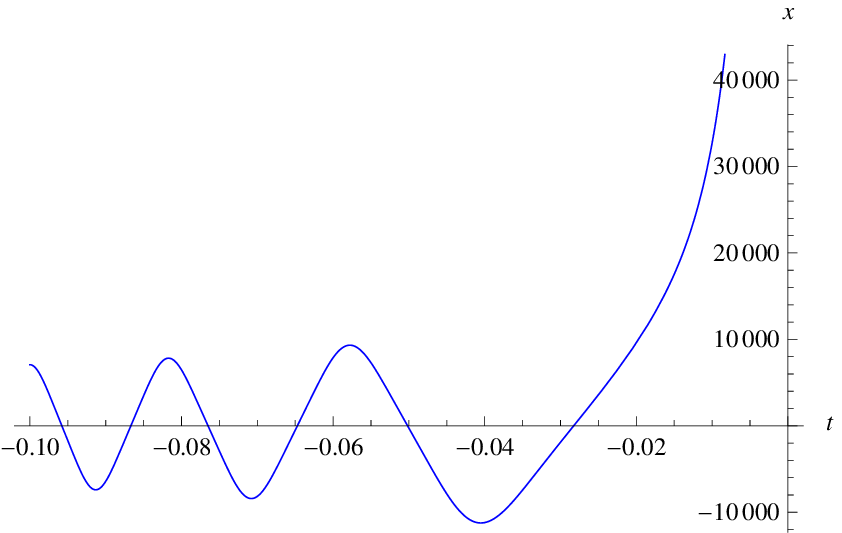}}
\mbox{\includegraphics[width=6cm]{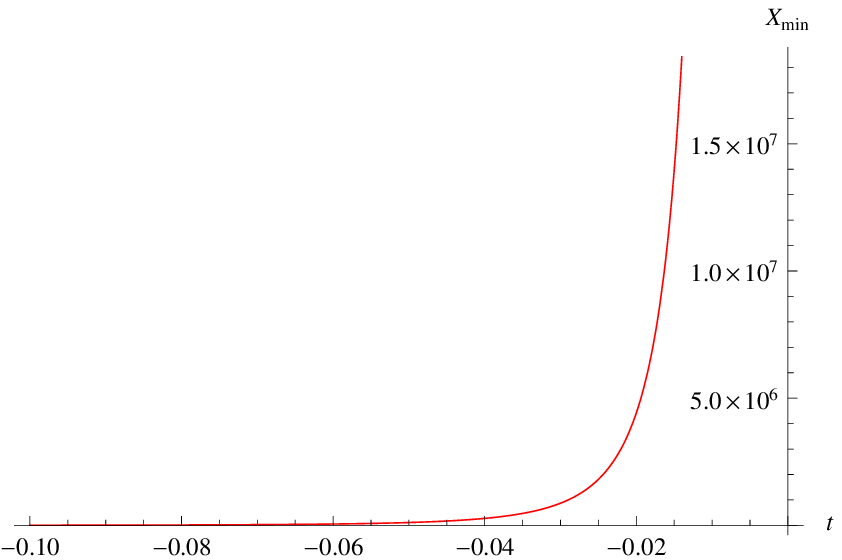}}}
\caption{Numerical solution for the motion with $p=2$, $q=3$ and the particle
at rest in the minimum of the potential at $t_0=-0.1$. The left plot is the
motion of the particle and the right plot the motion of the minimum.}
\end{figure}

As a consequence of these considerations we can make the strong
statement: \emph{for $t$ close (enough) to the singularity the
  evolution is always driven by the divergent tachyonic mass}.

In section 4 we will find that much of the behaviour revealed in this
classical analysis can be extended and adapted also to the quantum
mechanical evolution when we ``smear'' the particle into a wave
function.

\subsection{Classical analysis: late times}

Before turning to the quantum mechanics and 
to develop a qualitative understanding of the role of the quartic term
for $|t|\to\infty$ we will perform a change of variables by
multiplying the matrix coordinates by a power of $t$ so as to simplify
the $t$ dependence,
\begin{equation}
X = t^\gamma W,\,\, Y=t^\gamma Z.
\end{equation}

In addition we choose a new time parameter 
$$T \sim t^{-2\gamma +1}$$ where $\gamma = -q/3$ and for $q>0$
small/large $t$ corresponds to small/large $T$.

At large $T$ the leading terms in the action of our toy model \eqref{stoy} 
become
\begin{multline}
S_\mathtt{toy}=\frac{1}{2}\!\int\! dT\, \mathtt{Tr} 
\left( \left|\frac{dW}{dT}\right|^2\right. 
\!\!\!+\left|\frac{dZ}{dT}\right|^2 
\!\!+ m^2(|W|^2 + |Z|^2)T^{-4q/(2q+3)}\\ 
   -\left. 
(4q(2q+3)-p(p+2))\frac{(|W|^2 + |Z|^2)}{36T^2} 
- \lambda |[W,Z]|^2\right).
\end{multline}
$m^2$ in this equation contains the constant parts of the frequencies
(taken to be equal for simplicity) of the two $SU(2)$ coordinates, $X$
and $Y$.  Given that $q>0$ the time dependence in the $m^2$ term for
large $T$ is subleading when compared to the quartic term but always
more significant than the quadratic $1/T^2$ term.  For large $T$ we
are more interested in the behaviour of coordinates with $m^2=0$ as
only these have the possibility of escaping to large coordinates and
commuting configurations that are indicative of the emergence of
string theory in flat space-time.  This and related systems have been
much studied in the context of hyperbolic billiards and the mixmaster
universe and have been shown to be classically chaotic
\cite{Biro:1994bk}. For large $T$ one might think that even though $W$
and $Z$ are chaotic, $X$ and $Y$ will go to zero due to the leading
power of $t$ in the coordinate transformation. This is not entirely
true as the chaotic dynamics implies that occasionally $W$ and $Z$
will travel a long way along one of the (infinite number of) valleys of the
potential and there is no obvious limit on the magnitude of these
excursions. These general considerations are in complete agreement
with the large $|t|$ numerical simulation that were presented in
section 3.2.

It is notable that there is no obvious way in which the quartic term
is forcing the matrices to commute even though standard reasoning
would lead one to such a conclusion. The coefficient of the quartic
term in the toy model is indeed divergent at large $T$ indicating that
energetically favoured trajectories correspond to commuting matrices.
The classical evolution from strong to weak string coupling
however does not force the system into such a configuration. This result
presents us with two difficulties that we will revisit when we turn to
the quantum mechanics of this system. The first difficulty we have is
the apparent non-recovery of commuting space-time physics as one would
expect for the weakly coupled string theory in an asymptotically flat
space-time. The second problem is that, even if we could prove such
commutativity, we still have no control over the precise direction in the
configuration space where this commutativity will arise.  In quantum
mechanics this means that we could imagine a wave-function in which
several different possibilities for commuting matrices are present but
with these different possibilities not mutually commuting. At the end
of the next section we will turn to look at the large $t$ quantum mechanics.

\section{The quantum physics}

We will now move on to the quantum mechanics and in particular to the
examination of the applicability of time-dependent perturbation theory
to the quartic potential as a perturbation of the time-dependent
harmonic oscillator.  We begin our study of the evolution of the
quantum mechanical system for $|t|$ close to zero.

\subsection{The time-dependent harmonic oscillator}

In this first subsection we will present the solution to the
time-dependent harmonic oscillator further streamlining \cite{tdnotes} 
the alternative
presentation in \cite{Blau:2002js} rather than following the original
work of Lewis and Riesenfeld \cite{Lewis:1968yx,Lewis:1968tm}.
Recalling the solutions for the time-dependent harmonic oscillator
discussed at the beginning and promoting them to solutions to the
Heisenberg equations of motion for the operator $\hat{X}$ gives
\begin{equation}
\hat{X}(t)=\hat{a}F(t)+\hat{a}^\dagger F^{*}(t),
\end{equation}
where
\be
F(t) = \sqrt{\omega t}H_\nu(\omega t)
\ee
and the $F(t)$ are normalised such that the Wronskian $W(F,F^*)=i$.

The time-independent operators are creation and annihilation operators and 
can be seen to correspond to a choice of vacuum at infinity as a consequence
of the asymptotics of the Hankel function at $x\to\infty$,
\begin{equation*}
H_\nu(x) \approx \sqrt{\frac{1}{2\pi x}}e^{i(x-\frac{\pi\nu}{2})}.
\end{equation*}

The raising and lowering operators in $\hat{X}$ are by construction
time-independent and thus invariant,
\be
i\frac{d\hat{a}}{dt} =0
\ee
naturally leading to the definition of an invariant Hermitian operator 
\be
\hat{I} = \hat{a}^\dagger\hat{a} + \frac{1}{2},
\ee
\begin{equation}
\frac{d\hat{I}}{dt} = \frac{\partial\hat{I}}{\partial
  t}+\frac{1}{i\hbar}[\hat{I},\hat{H}]=0.
\end{equation}

Eigenvectors and eigenvalues of $\hat{I}$ are built using $\hat{a}$
and $\hat{a}^\dagger$ exactly as one does for the time-independent
harmonic oscillator leading to
\begin{equation}\label{oscbasis}
\begin{split}
\hat{a}^\dag \hat{a}|s\rangle&=s|s\rangle,\\
\hat{a}|s\rangle&=s^{\frac{1}{2}}|s-1\rangle,\\
\hat{a}^\dag|s\rangle&=(s+1)^{\frac{1}{2}}|s+1\rangle.\\
\end{split}
\end{equation}

The wave-function $\psi_0(x) = \langle x|0\rangle$ can then be found by 
solving the equation
\be
\langle x|\hat{a}|0\rangle = 0
\ee
and is (up to some factors of $\pi$ etc..)
\be
\psi_0(x,t) = \frac{1}{|F|^{1/2}} e^{ix^2\dot{F}^*/2F^*}.
\ee

Using the time-independence of the oscillators and taking a partial
time derivative of $\hat{a}^\dagger|0\rangle$ one finds that acting
with the Schr\"odinger operator on the vacuum gives
\be 
(i\partial_t - H)|0\rangle = \varphi(t)|0\rangle .
\ee 
To obtain an
explicit expression for $\varphi(t)$ we now multiply this equation on
the left by a position eigenstate $\langle x|$ and insert the above
ground state wave-function. Thus 
\be 
\varphi(t) =
-\frac{i}{2|F|}\frac{d|F|}{dt} +i \frac{\dot{F}^*}{2F^*} 
\ee 
and a simple calculation gives 
\be 
\varphi(t) = -\frac{1}{4|F|^2}.  
\ee

Extending this calculation to a general eigenstate we see that 
\begin{equation}
(i\partial_t -
\hat{H})|s\rangle =\varphi(t)|s\rangle,
\end{equation}
and as a consequence $\varphi(t)$ is independent of the
state $|s\rangle$. Thus the state vectors $|\tilde{s}\rangle$ are 
constructed from the corresponding $|s\rangle$ as
\begin{equation}\label{statevec}
|\tilde{s}\rangle = e^{-i\varphi(t)}|s\rangle.
\end{equation}
The behaviour of the phase $\varphi(t)$ as $t$ approaches the origin
can easily be seen to be 
\be
\varphi(t) \sim t^{p+1}
\ee
and it is straightforward to see that this phase does not 
exhibit any singularity at this point.
However, if we happen to choose to project these eigenstates onto a position 
basis then one does find a singularity. Indeed
\begin{equation}\label{wvfn}
\psi(x,t)=\langle x,t|s\rangle=N|F|^{-1/2}\mbox{e}^{ix^2\dot{F}^*/2F^*}
\mbox{e}^{-i\varphi(t)}H_s\left(\frac{x}{\hbar|F|}\right)
\end{equation}
where the $H_s$ are Hermite polynomials. Note that this wave-function
is not well defined when the singularity is approached due to the
phase, proportional to $\dot{F}^*/F^*$, which changes rapidly as $t$
approaches zero \cite{Awad:2008jf}. On the other hand, the state
vectors \eqref{statevec} that are solutions to the Schr\"odinger
equation are well-behaved as $t$ approaches zero and can be continued
from positive to negative $t$.

As is adequately discussed by Lewis and Riesenfeld \cite{Lewis:1968tm}
these state-vectors allow the extension of probability transitions
through t=0.  Thus, despite the fact that in the coordinate
representation one needs to deal with diverging coordinates and
rapidly varying phases there is nevertheless a way of describing
evolution through the singularity. Indeed if we assume that
\eqref{time-dep-freq} holds for all non-zero times (so that the mode
has the same asymptotic frequency $\omega_n$ for
$t\rightarrow\pm\infty$) we can choose as two independent real
solutions to \eqref{modeeq} $\sqrt{\omega_n t}J_\nu(\omega_n t)$ and
$\sqrt{\omega_n t}N_\nu(\omega_n t)$ for all $t\neq 0$. In
\cite{Lewis:1968tm} it is then explained how to use these solutions to
find the transition amplitude between eigenstates of the asymptottic
initial and final oscillators; in our case, as the asymptotic
frequencies are the same, we obtain the trivial result
\begin{equation}
T_{nm} = \langle n,t=\infty|m,t=-\infty\rangle = \delta_{nm}\;.
\end{equation}

Defining the transition through the singularity 
however remains an important problem as in principle one could
choose to glue modes with different asymptotic frequencies at
$t=0$. In addition some extra information on the behaviour of
$\hat{X}$ on both sides of the singularity is needed in order to
compute the amplitudes. Finally, the extension of this discussion to
QFT forces one to confront additional questions related to particle
creation and the choice of vacuum. These questions will be addressed
in a forthcoming publication \cite{tocome}.

\subsection{Time-dependent perturbation theory}

We will briefly look at the possibility of treating the quartic term as
a perturbation by using time-dependent perturbation theory. The
unperturbed states will be given by the solutions to the Schr\"odinger 
equation as provided in the previous subsection.
Consider the following Hamiltonian,
\begin{equation}
H = H_0(t) + \lambda \hat{W}(t)
\end{equation}
where $H_0(t)$ is the Hamiltonian for the time-dependent oscillator
while $W(t)$ is the quartic commutator term
\begin{equation}
\hat{W}(t) = -|t|^{2q}\mathtt{Tr}[\hat{X},\hat{Y}]^2.
\end{equation}

We will expand our full wave-function as a series with time-dependent
coefficients
\begin{equation}
|t\rangle = \sum_s b_s(t) e^{i\varphi(t)}|s\rangle.
\end{equation}
The requirement that this state is a solution to the full Schr\"odinger
equation leads to the first order differential equation for $b_s(t)$
\begin{equation}
\frac{db_s(t)}{dt} = \frac{\lambda}{i\hbar} \sum_{s^\prime} \langle
s|\hat{W}|s^\prime\rangle b_{s^\prime}(t).
\end{equation} 

To find a perturbative solution to this equation we now also expand
\begin{equation}
b_s(t) = b_s^{(0)} + \lambda b_s^{(1)}(t) + \lambda^2 b_s^{(2)}(t) + \cdots
\end{equation}
and retain the first non-trivial order. Specializing these general
equations to our case, we have to keep in mind that the non-Abelian
interaction couples six different copies of our time-dependent
harmonic oscillator (with asymptotic frequencies possibly different),
so if we represent the tensor product of six $\hat{I}$ eigenstates by
\begin{equation*}
|s_M\rangle=|s_{X1},\dots,s_{Y1},\dots\rangle
\end{equation*}
and we decompose the operators
$\hat{X}$ and $\hat{Y}$ then the perturbation has the general form
\begin{multline} 
\hat{W}\sim-|t|^{2q}\sum_{i,j,l,m}F_{\omega_{X_i}}F_{\omega_{Y_i}}
F_{\omega_{X_l}}F_{\omega_{Y_m}}(\delta_{il}\delta_{jm}-\delta_{ij}\delta_{lm})\times\\
\times\,(a_{X_i}+a_{X_i}^\dagger)(a_{Y_i}+a_{Y_i}^\dagger)(a_{X_l}+a_{X_l}^\dagger)
(a_{Y_m}+a_{Y_m}^\dagger).
\end{multline} 
The differential equation that determines $b_{s_M}^{(1)}(t)$ is then
\begin{equation}
\frac{db^{(1)}_{s_M}(t)}{dt} = \frac{\lambda}{i\hbar} 
\sum_{s_M^\prime}\langle s_M|\hat{W}| s_M^\prime\rangle b_{s_M^\prime}(t)
\end{equation}
where the sum is performed over all combinations
$s_M\!=\!\{s_{X1},\dots, s_{Y1}, \dots\}$. To find
$b^{(1)}_{s_M}(t)$ we need to solve a combinatorial problem and carry
out an integration over $t$. For our discussion it is sufficient to
consider the integration without elaborating on the details of the
combinatorics\footnote{As we are interested in the limit as $t\rightarrow 0$,
the appearance of different asymptotic frequencies is not important. For 
the same reason we have taken the lower limit of the integration to be 
$t=-\infty$.}
\begin{equation}
b^{(1)}_{s_M}(t) \sim \int_{-\infty}^t dt^\prime|t^\prime|^{2q}|F(t^\prime)|^4.
\end{equation}

We can investigate the convergence of this integral for $t$ near zero
by expanding the Bessel functions in $|F(t)|$ around $t=0$. We find that
$|F(t)|^2 \sim |t|^{-p}$ resulting in 
\begin{equation}
b^{(1)}_{s_M}(t) \sim \int^{t} |t^\prime|^{2(q-p)} \sim |t|^{2(q-p)+1}
\end{equation}
unless $q=p-1/2$ for which case the integral has a logarithmic divergence. 
Thus $b^{(1)}_s(t)$ converges as $t\rightarrow 0$ for $2(q-p)>-1$. Going back 
to the two cases that we have been discussing throughout the paper we see
that when $q=p(p+2)/4$ ($d=6$) we would require that $p(p-2)>-2$ 
which is always satisfied for 
$p>0$ and similarly so for other values of $d$. On the other hand, 
when $p=2q$ perturbation theory is only valid in the range 
$0<p<1$. In the next section we will investigate in greater
depth the apparent breakdown of perturbation theory when $p>1$, and we will
show that it is due to transient behaviour related to wave-functions that are
large in the region where the quartic term becomes significant.

\subsection{Evolution of wave-functions}

To find the source of the problems that arose in the previous section
when applying time-dependent perturbation theory to the case with
$p=2q$ and $p>1$ we begin with an investigation of the time-evolution
of the solutions to the Schr\"odinger equation in the inverted
harmonic oscillator and we compare their width and the rate at which
the wave-functions spread in this potential to the movement of the
location of the minimum in the full oscillator plus quartic
potential. We will once again simplify to a one-dimensional anharmonic
oscillator \eqref{onedim} with the time-dependence of our DLCQ model.
The action is
\begin{equation}
S =\frac{1}{2}\int\,dt\bigl(\dot{X}^2-\omega(t)^2 X^2-\lambda|t|^{p} 
X^4\bigr).
\end{equation}
We will use the Ehrenfest theorem to study the general features of
the evolution of the wave-function. As is well known, the Ehrenfest theorem is
exact for an harmonic oscillator potential and it can easily be seen
that this exactness remains when extended to a time-dependent
oscillator.

Corrections to the Ehrenfest theorem may arise when the width of the
wave-packet becomes comparable to the curvature of the potential as in
that case one can no longer use the approximation
\begin{equation}\label{Ehrapprox}
\left\langle \frac{dV(X)}{dX}\right\rangle \cong 
\frac{dV(\langle X\rangle)}{dX},
\end{equation}
that plays an important part in the derivation of the theorem.
Making a Taylor series expansion of $V(X)$ around $X=\langle X\rangle$ 
we can find the corrections to \eqref{Ehrapprox}
\begin{equation}\label{ehr}
\left\langle \frac{dV(X)}{dX}\right\rangle \cong
\frac{dV(\langle X\rangle)}{dX}
+ \frac{1}{2}\frac{d^3V(\langle X\rangle)}{dX^3}\langle(\Delta X)^2\rangle,
\end{equation}
and it is clear that for a purely quadratic potential there are
actually no such corrections.  To determine the full consistency of
this analysis one needs to consider that for our quartic potential the
third derivative term is no longer zero. In the Heisenberg
representation fluctuations of $\hat{X}$ also obey \eqref{pert2}
and so we can say that the corrections coming from the quartic term are
subleading as $t\to 0$. It is then obvious that its expectation value
$\langle X\rangle$ in any state also behaves asymptotically as if
there was no $X^4$ term in the potential. Moreover it is easy to show
that for the inverse harmonic oscillator $\langle(\Delta X)^2\rangle$
has at most the same divergence as $\langle X\rangle^2$, so that the
second term in the rhs of \eqref{ehr} is subleading: this can easily
be found by solving the differential equations for the expectation
values $\langle X^2\rangle$, $\langle P^2\rangle$.\footnote{Obviously 
these equations
  will be modified by the $X^4$ term, but the extra contributions are
  subleading as $t\to0$.} Thus, even if we include the cubic term
to investigate corrections to the Ehrenfest theorem, we find that for
$t$ sufficiently small these corrections are irrelevant.

Obviously if the initial wave-function has a significant non-zero
support in the region of the quartic minimum its initial evolution
will be significantly different from that of the inverted harmonic
oscillator. However, by considering the behaviour of $\langle(\Delta
X)^2\rangle$ and $\langle X\rangle^2$ as $t\rightarrow 0$ we conclude
that after a certain period of time the wave-function no longer has
significant support in the region of the minimum and its evolution is
precisely that of the inverted oscillator for which the Ehrenfest
Theorem applies. During this initial transient period the
wave-function will have acquired a modified profile but it is
guaranteed that after this period has passed the further evolution is
determined precisely by the inverted harmonic oscillator potential.

To understand this evolution in a little more detail 
we consider $p>1$ and take a
Gaussian wave-packet centred on $X=X_0$ as the initial condition at
$t=t_0>0$. This wave-packet can be decomposed into a sum over the
$\hat{I}$ eigenfunctions evaluated at $t=t_0$. This decomposition allows
us to examine in detail the time evolution of the wave-packet as $t
\rightarrow 0$. When the Gaussian is centred on $X_0<X_{\mathrm{min}}$
and its width at $t=t_0$ is much smaller than the distance to the
minimum of the full potential we find that this wave-function
spreads more slowly than the movement of the minimum (which obviously
goes to $\infty$ as $t\rightarrow 0$). The same behaviour is also
found for initial conditions on the Gaussian wave-packet such that its
width is not necessarily small but is however less than the distance
to the minimum, essentially requiring that the probability of finding
the particle described by this wave-function in the minimum of the
potential is very small.

The actual feasability of the numerical calculations that we performed
depends in an important way on the initial conditions. Indeed we
managed to carry out reliable numerical computations only for the case
of a very thin initial packet centred on $x=0$, while for a more
spread out initial wave-profile or one centred near the minimum of the
potential a sufficiently accurate truncation of the series requires
too many levels making a numerical evaluation of the evolution
impractical.  This notwithstanding the structure of the $\hat{I}$
eigenstates makes us confident enough to say that the spreading of all
wave-functions goes roughly as $|F(t)|$ as one can immediately see from
\eqref{wvfn}, namely like $t^{-\frac{p}{2}}$, while the minimum receeds to
infinity like $t^{-\frac{p}{2}-1}$. The inner-product of the
$\hat{I}$ eigenfunctions $\psi_l$ is obviously time-independent;
moreover considering the integral of $\psi_l\psi^*_{l'}$ from 0 to
$\infty$ one finds that this ``half product'' is also finite as $t\to
0$.  This means that all the states are growing more or less with the
same speed, and if one of them is losing contact with the minimum, the
others will lose it as well. In the previous section we found a
breakdown in the naive construction of time-dependent perturbation
theory for $p>1$ as a consequence of the growth with $p$ of the
characteristic width of the eigenfunctions of the Schr\"odinger
equation. We have now verified that this does not imply that given a
set of initial conditions and for $t$ sufficiently close to $0$ one
cannot however use time-dependent perturbation theory to describe the
evolution once the transient has passed.

We have thus reached the conclusion that for $p=2q>1$ and for an
arbitrary initial wave-function profile at some value $t=t_0$, the
evolution towards the singularity will consist of an initial transient
behaviour during which the non-linear effects of the quartic
interaction are important, followed by an evolution for which the only
important part of the potential is the purely quadratic inverted
harmonic oscillator; thus even in this case, after this initial
transient period, the quantum mechanics reduces to that of the
time-dependent harmonic oscillator without the quartic interaction
term.

\subsection{$|t|\to\infty$}

As mentioned at the conclusion of our discussion of the classical
physics of our toy model there are some surprises in the behaviour of
the system in the limit $|t|\to\infty$. These continue to be present
after quantisation.  A complete study of the quantum mechanics
requires a deeper understanding of the evolution of the wave-function
and thus a very careful numerical study of the Schr\"odinger equation
with time-dependent mass term and time-dependent potential.

In the supersymmetric case it is believed that the effective potential
is truly flat for commuting configurations \cite{Witten:1995im}
leading to the emergence of weakly coupled string theory in flat
space-time and this should continue to be true for the time-dependent
system that we are studying. For the analogous
mixmaster universe, where our evolution to large $t$ corresponds
mathematically to the evolution towards the mixmaster singularity,
some calculations of the evolution of wave-packets have been carried
out by Furusawa \cite{Furusawa:1985ef,Furusawa:1986fg} and later also
by Berger \cite{Berger:1989jm}. Their results indicate that the system
remains chaotic upon quantisation and that an initially smooth
wave-function will break up into many differently localised
fragments. This leads one to conclude, as we already argued in the
discussion of the classical physics, that the system does not settle
down into a fixed commuting configuration at late times but rather
explores all different possible commuting configurations as
$|t|\to\infty$.

\section{Summary}

As already emphasized in the introduction, the work presented in this
paper was motivated by the desire to understand, in as thorough a
manner as possible, the time-dependent gauge theory that one finds
when the DLCQ procedure is generalised to SHPWs
\cite{Blau:2008bp}. This time-dependent Yang-Mills theory presents
interesting features arising from the interaction
between time-dependent frequencies and couplings on the one hand and
non-Abelian matrix fields on the other. In particular we have addressed 
two specific questions: To what extent is non-commutativity of the 
matrix fields relevant near the singularity? and: How does
an intrinsically non-commutative system of matrices at $t=0$ evolve into
a system of commuting coordinates as $t\rightarrow\infty$? Obviously the
answers to these questions are also of interest for the study of
time-dependent gauge theories independently of their matrix theory
origins.

We have approached these questions at both the classical and quantum
mechanical level, leaving an extension of this discussion to quantum
field theory for a forthcoming paper \cite{tocome}. 
The classical analysis revealed two interesting facts:
\begin{enumerate}
\item
It is often said that near the singularity the non-commutativity of
the non-Abelian degrees of freedom should play an important role.
This is certainly true in the sense that for $|t|$ small the
off-diagonal degrees of freedom are light and can easily be excited,
but this does not mean that the explcitly non-Abelian quartic
interaction drives the evolution. On the contrary the
singular quadratic term (``divergent tachyonic mass'') eventually
dominates in all cases and the non-Abelian interaction can be treated
as a perturbation that is irrelevant at small times. 
\item
For large times the system fails to settle on a specific commuting
arrangement of coordinates and appears to move randomly between
different commuting configurations. In some sense the geometry can
still be said to be commuting, as the coordinates spend most of their
time in nearly commuting configurations and the transitions between
different configurations are fast.  
\end{enumerate}   

We began the quantum mechanical analysis with a new discussion of the
solution to the time-dependent harmonic oscillator \cite{Lewis:1968yx}
anticipated in \cite{Blau:2002js} and completed here.  We then
demonstrated that the quartic interaction is subleading as $t\to 0$
also in quantum mechanics. If one relies on the simplest analysis of
perturbation theory this result appears to not hold in some of the
cases that we studied but this is simply due to a non-linear transient
phase of the evolution during which the interaction is not necessarily
negligible. We show in generality that in all cases of interest the
mass term eventually dominates the potential, a result we also
anticipated in numerical calculations.

In general we have learnt that, even if the physics near the
singularity is intrinsically non-Abelian as far as access to
off-diagonal degrees of freedom is concerned, the non-Abelian
interaction actually becomes less and less important as we approach
$t=0$. What remains to be done is: the investigation of the 
extension of these results to quantum field theory; 
an analysis of the transient period during which there is also a radical 
change in the number of accessible degrees of freedom; and a definite
interpretation of what happens at $t=0$ and as $t\rightarrow\infty$
\cite{tocome}.

Finally, an intriguing question that has been part of our continuing motivation 
for studying these matrix theories is the universality of the singularity 
and the simple scaling of the matrix actions under rescalings of $t$ near 
$t=0$. Unfortunately we do not have anything new to add to the understanding
of this symmetry but we hope that our demonstration in this article of a 
further universal role for the divergent tachyonic masses
in the near singularity 
region will provide additional impetus for further study of this question. 

\vskip 1cm
\noindent
{\bf Acknowledgements}

The authors would like to thank Matthias Blau and Darko Veberi\v c 
for various helpful discussions.

\bibliographystyle{unsrt}
\end{document}